\documentclass[11pt]{article}
\usepackage{graphicx}
\usepackage{color}

\usepackage[english]{babel}
\usepackage{pgf,pgfarrows,pgfnodes,pgfautomata,pgfheaps}
\usepackage{amsmath,amssymb,exscale}
\usepackage[latin1]{inputenc}
\usepackage{eucal}
\usepackage{textpos}
\usepackage{epsfig}
\usepackage{color}
\usepackage{mathrsfs}
\usepackage{multicol}
\usepackage{blindtext}
\usepackage{mathtools}
 \usepackage{fancyhdr}
\usepackage{graphicx}
\usepackage{hyperref}
\usepackage{hyperref}

\newcommand\pubnumber{CERN-TH-2017-243}
\newcommand\pubdate{\today}

\newcommand{\GeV}{\,{\rm GeV}}
\newcommand{\bp}{\bar{M}_{\rm Pl}}

\def\institute{Theoretical Physics Department\\
CERN,  Geneva, Switzerland}
\def\support{
         }

\def\Title#1{\begin{center} {\Large #1 } \end{center}}
\def\Author#1{\begin{center}{ \sc #1} \end{center}}
\def\Address#1{\begin{center}{ \it #1} \end{center}}

\newcommand\pubblock{\rightline{\begin{tabular}{l} \pubnumber\\
         \pubdate  \end{tabular}}}
\newenvironment{Abstract}{\begin{quotation}  }{\end{quotation}}
\newenvironment{Presented}{\begin{quotation} \begin{center} 
             PRESENTED AT\end{center}\bigskip 
      \begin{center}\begin{large}}{\end{large}\end{center} \end{quotation}}
\def\Acknowledgements{\bigskip  \bigskip \begin{center} \begin{large}
             \bf ACKNOWLEDGEMENTS \end{large}\end{center}}

\def\beq{\begin{equation}}
\def\eeq#1{\label{#1}\end{equation}}
\def\eeqn{\end{equation}}

\def\beqa{\begin{eqnarray}}
\def\eeqa#1{\label{#1}\end{eqnarray}}
\def\eeqan{\end{eqnarray}}

\let\bar=\overbar

\def\Dslash{\not{\hbox{\kern-4pt $D$}}}
\def\dslash{\not{\hbox{\kern-2pt $\del$}}}

\def\msb{{\bar{\ssstyle M \kern -1pt S}}}

\begin{document}
\begin{titlepage}
\pubblock

\vfill
\Title{The Electroweak Vacuum Decay \\ and the Gravitational Contribution}
\vfill
\Author{ Alberto Salvio\support}
\Address{\institute}
\vfill
\begin{Abstract}
\noindent Whether the Standard Model electroweak vacuum is stable, metastable or unstable depends crucially on the top mass (and, to a lesser extent, on other measurable quantities). These topics are reviewed and updated by taking into account the most recent determination of the top and Higgs masses. Moreover, the correction to the vacuum decay due to Einstein gravity  is described. This process is also discussed in a scenario, called softened gravity, which has been proposed as a solution of the Higgs mass hierarchy problem. 
\end{Abstract}
\vfill
\begin{Presented}
$10^{th}$ International Workshop on Top Quark Physics\\
Braga, Portugal,  September 17--22, 2017
\end{Presented}
\vfill
\end{titlepage}
\def\thefootnote{\fnsymbol{footnote}}
\setcounter{footnote}{0}

\section{Introduction}
Depending on the top mass $M_t$ and, to a lesser extent, to other observables such as the Higgs mass $M_h$ and the strong fine structure constants $\alpha_3$, three scenarios are possible in the {\it pure} Standard Model (SM). The electroweak (EW) vacuum expectation value $v$ can be stable, metastable (with a finite lifetime bigger than the age of the universe) or unstable. 
This paper is based on a talk I gave at the $10^{\rm th}$ edition of the top conference and I was, once again, glad to ascertain the interest  with which the community, including the experimentalists, look at this topic. 

Why should we pursue these studies? Let us recall that the main purposes of the LHC include testing the SM at the EW scale and discover new physics at those energies (if any!). Extrapolating the SM up to very high energies (even up to the Planck scale) provides a {\it complementary way to test the SM} as these energies cannot be probed at particle colliders. Therefore, these calculations find their best motivation when they are regarded as independent tests and they should not be interpreted as the commitment to the SM. Indeed, we know that  the SM has to be extended and the new physics, which is certainly hiding somewhere, may (or may not) change the results obtained in the pure SM. We will come back to this point in Sec.~\ref{noGrav}.

My talk was roughly divided {\it in three parts}. In the first one I reviewed the main results regarding the EW vacuum (meta)stability without gravity, on flat space-time. This introductory part was based on the state-of-the-art calculation of Ref.~\cite{Buttazzo:2013uya} and will be reported here in Sec.~\ref{noGrav}. In the second one, I presented recent calculations of the gravitational corrections to the vacuum decay based on the more recent paper~\cite{Salvio:2016mvj} and discussed here in Sec.~\ref{SoftSec}. Finally, I mentioned qualitatively the fate of the EW vacuum in a scenario motivated by the Higgs mass hierarchy problem: this problem can be solved, as I will review, if the growth of the Einstein gravitational interactions stops at some high energy scale no larger than $10^{11}\,$GeV~\cite{Giudice:2014tma,Salvio:2014soa,Salvio:2016vxi} {\it (softened gravity)}.  This final part is presented here in Sec.~\ref{SoftSec}. The various sections include the conclusions on the respective parts.

\section{The results without gravity} \label{noGrav}
\begin{figure}[t]
\centering
\includegraphics[height=2.in]{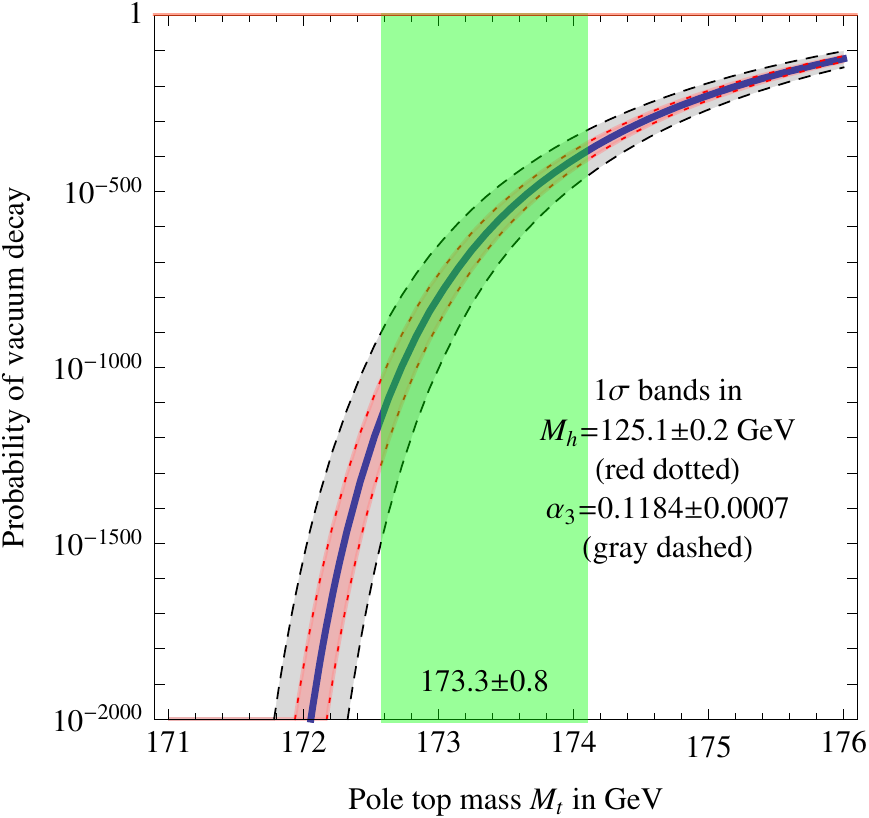} \qquad\qquad \includegraphics[height=2.in]{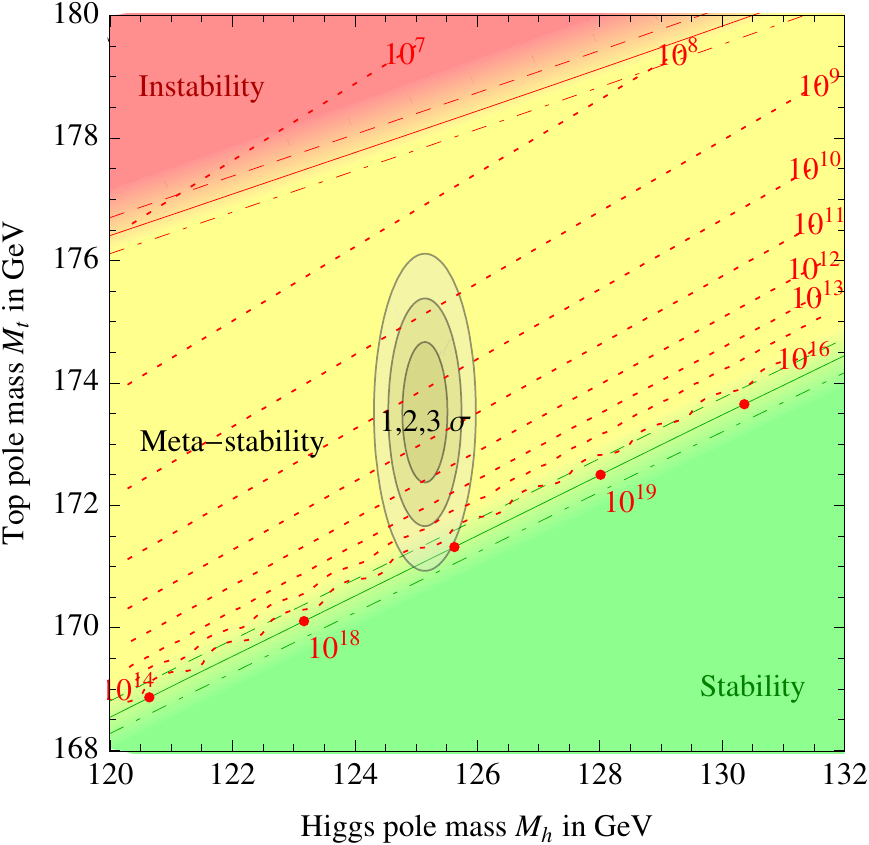} 
\caption{\it {\bf Left:} probability of vacuum decay; the  uncertainties due to $M_t$, $\alpha_3$ and $M_h$ are given in grey, red and blue respectively.  {\bf Right:}  SM Phase diagram without gravity.
Both plots are reproduced from \cite{Buttazzo:2013uya}.}
\label{fig:phase-prob}
\end{figure}

When $v$ is not stable,  the probability $d{\cal P}/dV\, dt$ per unit time  and volume of creating a  
bubble of true vacuum within a space volume $dV$ and time interval
$dt$ is given by
$
d{\cal P} =dt\,dV ~ \Lambda_B^4\, e^{-S_B}
$
\cite{Coleman}. Here 
$S_B$ is the action of the bounce of size $R\equiv \Lambda_B^{-1}$: the bounce $h(r)$  is an SO(4) symmetric solution of the Euclidean  Higgs equation with Higgs effective potential $V$,
\begin{equation} h'' + \frac{3}{r} h' = \frac{dV}{dh}, \quad \mbox{with boundary conditions} \quad h'(0)=0, \quad h(\infty) = v. \label{bounce-flat} \end{equation}
Physically, this represents the creation of a true-vacuum bubble  inside a false-vacuum environment so that the process resembles boiling water.  Even if this probability is non-zero, it turns out to be extremely small in the SM (see the plot on the left of Fig.~\ref{fig:phase-prob} and the precision calculations in) so that this worrisome fact is not actually harmful for the SM consistency\footnote{Whether the SM is compatible with the early universe data is another story though (see Refs.~\cite{Salvio:2013rja,1706.00792} for recent discussions)}

One question I have been asked during the conference is {\it ``why is the top mass so important for the stability of the SM?"} A qualitative answer can be easily provided. The top mass is approximately given by $M_t \approx y_t v$, where $y_t$ is the top Yukawa coupling. There are loop corrections to this expression, but one can neglect  them in a qualitative estimate as they are small. $y_t$ contributes negatively to the renormalization group (RG) $\beta$ function of  the quartic coupling of the Higgs, $\lambda$. The bigger $y_t$ is the more $\lambda$ can become negative at large RG energy $\mu$, destabilizing $v$ (see the left plot in Fig.~\ref{fig:lambda-grav} for the SM behavior). This leads to an upper bound on $M_t$. The currently most refined determination of this bound, known as the stability bound, is 
$$
M_t < (171.09\pm 0.15_{\rm th}\pm 0.25_{\alpha_3} \pm 0.12_{M_h})\GeV=
(171.09\pm 0.31)\GeV ,
$$
where the labels indicate the source of the corresponding uncertainties. The uncertainty $0.15_{\rm th}$ comes only from perturbative calculations of higher order than those considered in~\cite{Buttazzo:2013uya}. 
The bound above slightly differs from the one given in~\cite{Buttazzo:2013uya} because the more recent determination of $M_h$  has been used here~\cite{Aad:2015zhl}: 
$M_h = (125.09\pm  0.21_{\rm stat.} \pm 0.11_{\rm syst.})\GeV. $
See also Ref.~\cite{Andreassen:2017rzq}.
We observe thus some tension with the experimental values $M_t = 172.51 \pm 0.50\,$GeV (ATLAS) and $M_t=(172.44\pm 0.49)\,$GeV (CMS), which is, separately, at the 2-3$\, \sigma$ level. These new experimental values have been discussed at this conference~\cite{ATLAStop2017}. In Fig.~\ref{fig:phase-prob} (right) I give the SM phase diagram, where the ellipses correspond to the experimental uncertainties. The input parameters are those used in~\cite{Buttazzo:2013uya} (in particular $M_t$ was set to the world average $M_t =(173.76 \pm 0.76)\,$GeV plus an uncertainty of order $\Lambda_{\rm QCD} \approx 0.3\,$GeV).

As mentioned before these results can change completely (or remain the same) due to the new physics required by the experiments. For example, in~\cite{Salvio:2015cja} I provided a simple extension of the SM with right-handed neutrinos and an invisible axion sector, which is able to account for all observational hints of beyond the SM (neutrino oscillations, dark matter and baryon asymmetry), to solve the strong-CP problem, trigger inflation and stabilize the EW vacuum, even if $M_t$ is set to the current central value. On the other hand, the SM extended only by right-handed neutrinos  can account for
neutrino oscillations, dark matter and baryon asymmetry (see~\cite{Canetti:2012kh} and references therein) and predicts a  vacuum decay substantially identical to the pure SM one.




\section{Including Einstein gravity corrections}\label{EinsteinSec}

\begin{figure}[t]
 \begin{center}
  \includegraphics[width=2.1in]{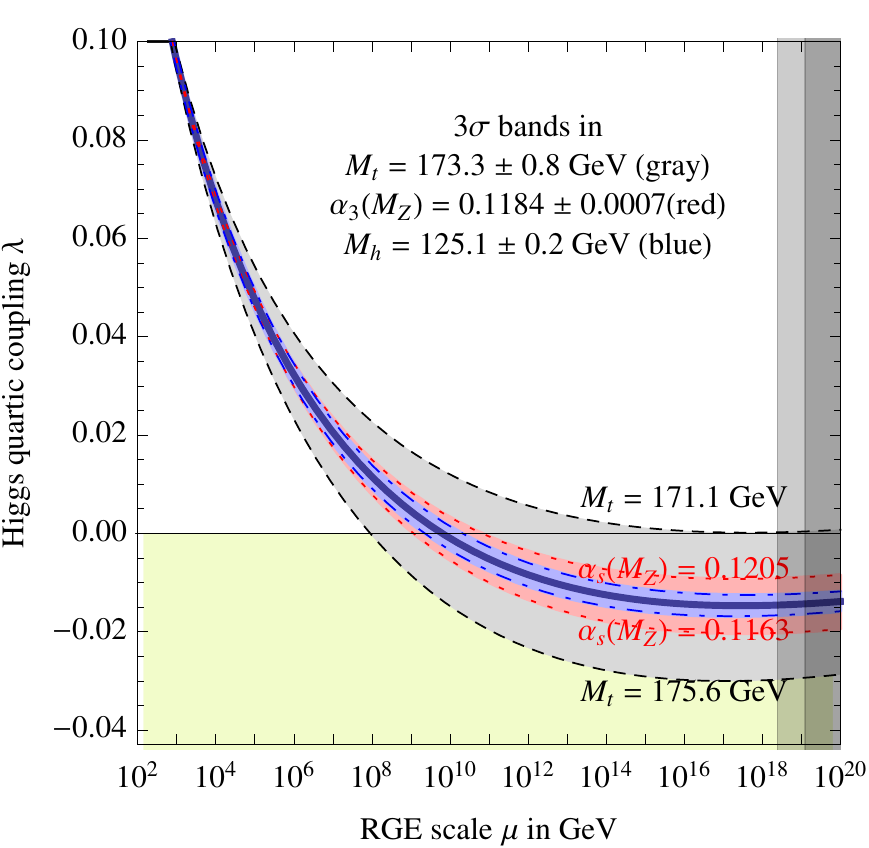} \qquad \qquad \includegraphics[width=2.1in]{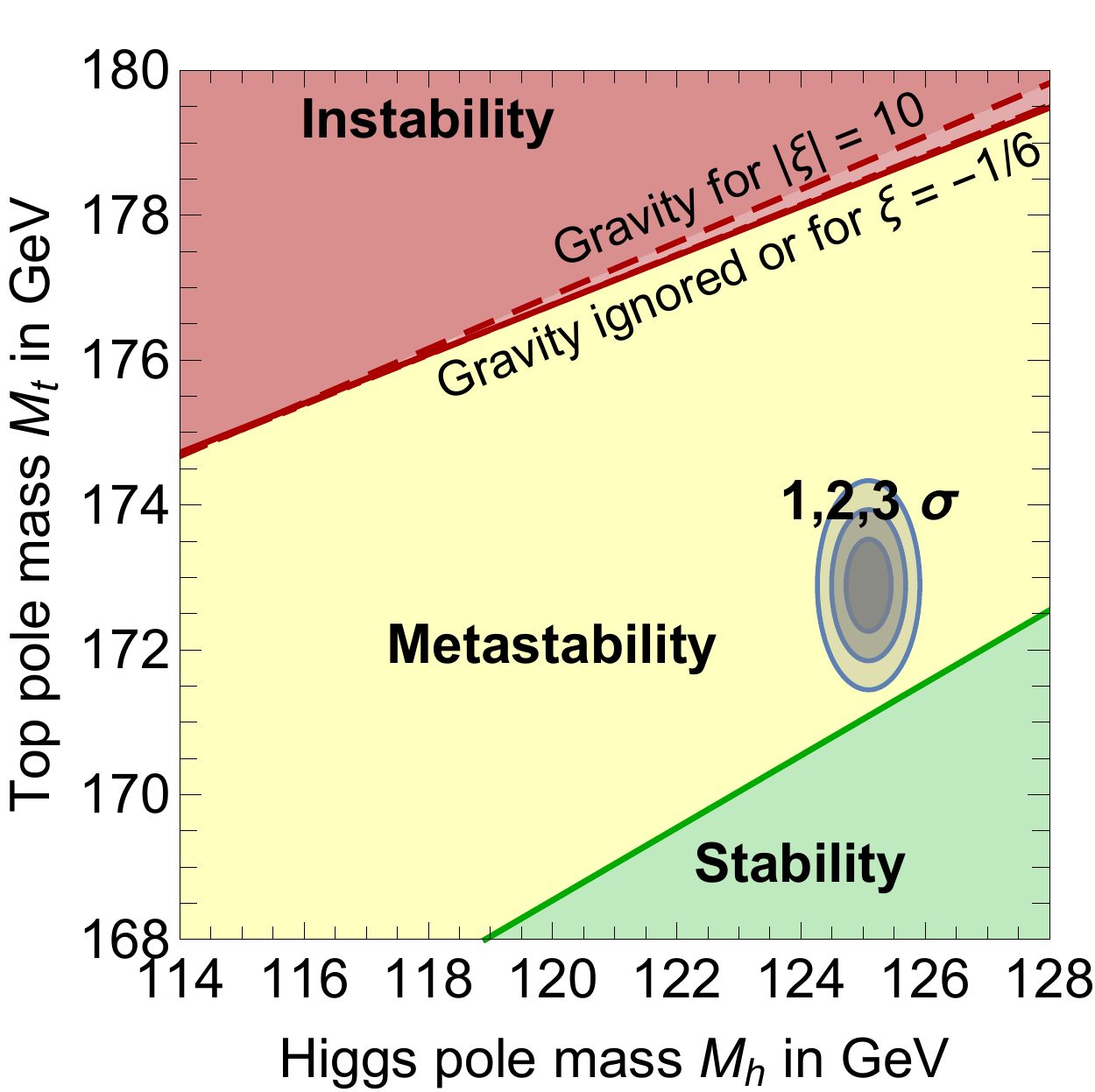}
\end{center}
\caption{\it {\bf Left:} Dependence of $\lambda$ on the renormalization group equation (RGE) scale. This plot is reproduced from \cite{Buttazzo:2013uya}. {\bf Right:} the SM phase diagram with Einstein gravity corrections. }
\label{fig:lambda-grav}
\end{figure}
We are looking at extremely high RG energies, sometimes reaching the Planck mass $M_{\rm Pl}$. 
 Does then gravity play a role? One can address this question in Einstein gravity,
  $$ \mathscr{L}  = \mathscr{L}_{\rm SM}+  \mathscr{L}_{\rm EH} 
- \xi |H|^2 \mathscr{R}, $$ which is compatible with all experiments.
Here $\mathscr{L}_{\rm SM}$ corresponds to the SM Lagrangian minimally coupled to gravity, $\mathscr{L}_{\rm EH}$ represents the standard Einstein-Hilbert Lagrangian for gravity and the last term is a non-minimal coupling between the Higgs doublet $H$ and the Ricci scalar $\mathscr{R}$, whose size is set by the coupling $\xi$. The latter term has been added because there is no rationale for excluding it from the Lagrangian\footnote{Higher dimensional terms such as $|H|^6$, etc. are instead expected to be suppressed at energies smaller than the Planck mass, and terms quadratic in the curvature, e.g. $\mathscr{R}^2$, add additional degrees of freedom besides those of the SM and a spin-2 massless graviton. We therefore ignore these additional terms for the sake of minimality.}. Here we are adopting an effective field theory approach as Einstein gravity breaks down at an energy around $M_{\rm Pl}$. It is therefore not restrictive to consider an expansion in $E/M_{\rm Pl}$, where $E$ is the typical energy of the process under study (for $E\gtrsim M_{\rm Pl}$ the theory anyhow breaks down).

Let us now give some details on how to technically include gravity. The bounce equation (\ref{bounce-flat}) becomes a Higgs-gravity system of equations
\begin{equation}
h'' + 3\frac{\rho'}{\rho} h' = \frac{dV}{dh} - \xi h  \mathscr{R},\qquad
\rho^{\prime 2} = 1 + \frac{\rho^2/\bp^2}{3 (1+ \xi h^2/\bp^2)}\left(\frac{h^{\prime 2}}{2} - V-6\frac{\rho'}{\rho} \xi h h'\right),\label{system-eq} \end{equation}
where $\bp \equiv M_{\rm Pl}/\sqrt{8\pi}$ is the reduced Planck scale and the function $\rho$ determines the metric of space-time $ds^2$ as follows:
\begin{equation}ds^2 = dr^2 + \rho(r)^2 d\Omega^2.\label{metric}\end{equation}
In the expression above $d\Omega$ is the volume element of the unit 3-sphere. The boundary conditions for $h$ at $r=0$ and $r=\infty$ remain the same as in Eq.~(\ref{bounce-flat}). Single-valuedness of the metric implies that $\rho(0)=0$: indeed, $r=0$ is a single point in the coordinates chosen in (\ref{metric}) and any value of the coordinates in $d\Omega^2$ should give the same metric at that point. In order to explain the boundary conditions for $\rho$ at large $r$ we have to specify the asymptotic space-time geometry far away from the bubble.   In this paper we will only consider  the case in which the false vacuum $v$ corresponds to the flat space-time (zero cosmological constant). Recently, the calculations presented in this paper have been generalised to the de Sitter space (positive cosmological constant) with interesting insights for  inflation~\cite{1706.00792}. I will not review this extension because of lack of space here. Therefore, given that the flat space-time has a $\rho$-function equal to $r$, we have to require $\rho(\infty)=r$. 

The system in (\ref{system-eq}) can be simplified by using perturbation theory in $1/(R\bp)$ without loss of generality. The inverse bounce size $1/R$ indeed represents the typical energy $E$ of the process under consideration and, as stated above, for $E$ bigger than the Planck scale Einstein gravity breaks down.

 The phase diagram of the SM including the Einstein gravity corrections is given in Fig.~\ref{fig:lambda-grav} (right)\footnote{I thank A. Urbano for quickly providing this plot during the conference.}. The gravitational corrections tend to stabilize the vacuum  in the asymptotically flat case under study, i.e. $\rho(\infty)=r$. This is not general: in the asymptotically de Sitter case the gravitational corrections could also have the opposite effect in some cases (see Ref.~\cite{1706.00792}). We observe that the gravitational corrections are quite small indicating that the result without gravity is a good approximation. In Fig.~\ref{fig:lambda-grav} (right) the value and uncertainty of the Higgs mass are those of Ref.~\cite{Aad:2015zhl}. Regarding $M_t$, a naive average of recent determinations~\cite{Sirunyan:2017idq} and the Tevatron result~\cite{CDF:2013jga} have been performed. This neglects the correlation between the measurements so it can be considered an optimistic estimate of the uncertainty (probably smaller than the actual one). A pessimistic estimate of the uncertainty (probably bigger than the actual one) is given by the old world average $M_t =(173.76 \pm 0.76)\,$GeV, used in Fig.~\ref{fig:phase-prob} (right). The conclusion is that the tension between the stability and the experimental inputs are at roughly the 3$\,\sigma$ level.

\section{Softened gravity and vacuum decay}\label{SoftSec}

In the SM coupled to Einstein gravity, analysed in the previous section, there is no explanation for the large hierarchy $v\ll \bp$. One way to address this hierarchy problem is to assume that the power-law increase of the Einstein gravitational coupling stops at $\Lambda_G \ll \bp$. The scale $\Lambda_G$ can be determined by noting  that the gravitational contribution to the Higgs mass is then 
$\delta M_h^2 \approx G_N \Lambda_G^4/(4\pi)^2$ at leading order in the Newton constant $G_N \equiv 1/(8\pi\bp^2)$. Indeed,  the effect should vanish as $G_N\rightarrow 0$ and the $1/(4\pi)^2$ suppression is due to the fact that this is a quantum correction, which appears only at loop level. Requiring then $\delta M_h^2 \lesssim  M_h^2$ (so that gravity does not give unnaturally large corrections to the Higgs mass) leads to the bound $\Lambda_G \lesssim  10^{11}\,$GeV, which has to be satisfied in order to solve the hierarchy problem. 

Given that gravity becomes soft at high energies it negligibly affects the stability issue.
In Ref.~\cite{Salvio:2016mvj} we have checked this conclusion in a concrete realization of softened gravity~\cite{Salvio:2014soa,Kannike:2015apa,Salvio:2016vxi}.

\Acknowledgements
I am grateful to the coauthors of Refs.~\cite{Buttazzo:2013uya,Salvio:2016mvj,Giudice:2014tma} for  inspiring discussions and collaborations related to the topics of this article. I also thank A. Castro, A. Giammanco and B. Pearson for useful discussions on the experimental determinations of the top mass. This work was supported by the grant 669668 -- NEO-NAT -- ERC-AdG-2014.

\end{document}